\documentclass[12pt]{article}
\usepackage{amsmath,amssymb,amsthm}

\newtheorem{theorem}{Theorem}[]
\newtheorem{lemma}{Lemma}

\newtheorem{claim}{Claim}
\newtheorem{definition}{Definition}

\newcommand{\join}{\text{\textcircled{{\footnotesize 1}}}}
\newcommand{\cojoin}{\text{\textcircled{{\footnotesize 0}}}}

\begin{document}

\title{Weighted Efficient Domination for $P_6$-Free Graphs}

\author{
Andreas Brandst\"adt\footnote{Fachbereich Informatik,
Universit\"at Rostock, A.-Einstein-Str. 21, D-18051 Rostock, Germany,
{\texttt ab@informatik.uni-rostock.de}}
\and
Raffaele Mosca\footnote{Dipartimento di Economia, Universit\'a degli Studi ``G. D'Annunzio''
Pescara 65121, Italy.
{\texttt r.mosca@unich.it}}
}

\maketitle

\begin{center}
{\em ad laudem Domini}
\end{center}

\begin{abstract}
In a finite undirected graph $G=(V,E)$, a vertex $v \in V$ {\em dominates} itself and its neighbors in $G$.
A vertex set $D \subseteq V$ is an {\em efficient dominating set} ({\em e.d.} for short) of $G$ if every $v \in V$ is dominated in $G$ by exactly one vertex of $D$. The {\em Efficient Domination} (ED) problem, which asks for the existence of an e.d. in $G$, is known to be NP-complete for $P_7$-free graphs but solvable in polynomial time for $P_5$-free graphs. The $P_6$-free case was the last open question for the complexity of ED on $F$-free graphs. 

Recently, Lokshtanov, Pilipczuk and van Leeuwen showed that weighted ED is solvable in polynomial time for $P_6$-free graphs, based on their sub-exponential algorithm for the Maximum Weight Independent Set problem for $P_6$-free graphs. Independently, at the same time, Mosca found a polynomial time algorithm for weighted ED on $P_6$-free graphs using a direct approach. In this paper, we describe the details of this approach which is simpler and much faster, namely its time bound is ${\cal O}(n^6 m)$.   
\end{abstract}


\section{Introduction}

Let $G=(V,E)$ be a finite undirected graph with $|V|=n$ and $|E|=m$. A vertex $v \in V$ {\em dominates} itself and its neighbors. A vertex subset $D \subseteq V$ is an {\em efficient dominating set} ({\em e.d.} for short) of $G$ if every vertex of $G$ is dominated by exactly one vertex in $D$.
The notion of efficient domination was introduced by Biggs \cite{Biggs1973} under the name {\em perfect code}.
Note that not every graph has an e.d.; the {\sc Efficient Dominating Set} (ED) problem asks for the existence of an e.d.\ in a given graph $G$.
If a vertex weight function $w: V \to \mathbb{N} \cup \{\infty\}$ is given, the {\sc Weighted Efficient Dominating Set} (WED) problem asks for a minimum weight e.d. in $G$, if there is one, or for determining that $G$ has no e.d. (instead of minimum weight one can ask for maximum weight as well; subsequently we restrict the problem to the minimum weight version).  
The vertex weight $\infty$ plays a special role; vertices which are definitely not in an e.d. $D$ get weight $\infty$, and thus, in the WED problem we are asking for an e.d. of  finite minimum weight. 

\medskip

The importance of the ED problem for graphs mostly results from the fact that ED for a graph $G$ is a special case of the {\sc Exact Cover} problem for hypergraphs (problem [SP2] of \cite{GarJoh1979}); ED is the Exact Cover problem for the closed neighborhood hypergraph of $G$.

\medskip

For a graph $F$, a graph $G$ is called {\em $F$-free} if $G$ contains no induced subgraph isomorphic to $F$.
Let $P_k$ denote a chordless path with $k$ vertices. $F+F'$ denotes the disjoint union of graphs $F$ and $F'$; for example, $2P_3$ denotes $P_3+P_3$.

Many papers have studied the complexity of ED on special graph classes - see e.g. \cite{BraEscFri2015,BraGia2014,BraMilNev2013,Milan2012} for references. In particular, a standard reduction from the Exact Cover problem shows that ED remains NP-complete for $2P_3$-free (and thus, for $P_7$-free) chordal graphs. 

\medskip

In this paper, we give a polynomial time solution for weighted ED on $P_6$-free graphs. For this graph class, the question whether ED can be solved in polynomial time was the last open case for $F$-free graphs \cite{BraGia2014}; it was the main open question in \cite{BraMilNev2013}.   
As a first step to a dichotomy, it was shown in \cite{BraEscFri2015} that for $P_6$-free chordal graphs, WED is solvable in polynomial time. 

\medskip

Recently, it has been shown by Daniel Lokshtanov, Marcin Pilipczuk and Erik Jan van Leeuwen  \cite{LokPilvan2015} that WED is solvable in polynomial time for $P_6$-free graphs in general; the result is based on their sub-exponential algorithm for the Maximum Weight Independent Set problem for $P_6$-free graphs (the time bound for WED is more than ${\cal O}(n^{500})$).

\medskip

Independently, based on the direct approach of Mosca, we obtain a ${\cal O}(n^6 m)$ time solution for WED on $P_6$-free graphs. Thus, the result of \cite{LokPilvan2015} and our approach finally lead to a dichotomy for the WED problem on $P_k$-free graphs and moreover on $F$-free graphs. In Section \ref{EDP6frpol}, we describe this direct approach which in a first step reduces WED on $P_6$-free graphs to the same problem for $P_6$-free unipolar graphs, and in a second step solves it for $P_6$-free unipolar graphs.




\section{WED on $P_6$-free graphs in polynomial time}\label{EDP6frpol}

Let $G=(V,E)$ be a finite undirected simple graph. For $U \subseteq V$ and $x \notin U$, we say that {\em $x$ contacts $U$} if $x$ has a neighbor in $U$, and {\em $x$ distinguishes $U$} if it has a neighbor and a non-neighbor in $U$. 
By $x \join U$ ($x \cojoin U$, respectively), we denote that for $x \notin U$, $x$ is adjacent to all vertices in $U$ ($x$ is non-adjacent to all vertices in $U$, respectively). If $x \join U$ ($x \cojoin U$, respectively), we say that {\em $x$ has a join to $U$} ({\em $x$ has a co-join to $U$}, respectively).  
If for $u \in U$, $u \join (U \setminus \{u\})$, we say that {\em $u$ is universal for $U$}.

\medskip

A graph $G=(V,E)$ is {\em unipolar} if there is a partition of $V$ into sets $A$ and $B$ such that $G[A]$ is $P_3$-free (i.e., the disjoint union of cliques) and $G[B]$ is a complete graph. See e.g. \cite{EscWan2011,McDYol2015} for recent work on unipolar graphs. Note that ED remains NP-complete for unipolar graphs \cite{EscWan2011} (which can also be seen by the standard reduction from Exact Cover; there, every clique in $G[A]$ has only two vertices).

\medskip

Our approach for solving the WED problem in polynomial time on $P_6$-free graphs is based on some properties of $P_6$-free graphs with e.d. 
In Subsection \ref{unipolarred}, we reduce the WED problem for $P_6$-free graphs to the same problem for $P_6$-free unipolar graphs and in Subsection \ref{sec:WEDP6frunipolargr}, we solve WED for such graphs in polynomial time. 

\medskip

Thus, we obtain a dichotomy for the WED problem on $P_k$-free graphs and on $P_k$-free unipolar graphs: For $k \ge 7$, WED is NP-complete for $P_k$-free unipolar graphs, and for $k \le 6$, WED is solvable in polynomial time for $P_k$-free unipolar graphs (clearly, any unipolar graph is $2P_3$-free and thus $P_7$-free).  

\subsection{Reducing WED on $P_6$-free graphs to $P_6$-free unipolar graphs}\label{unipolarred}

Let $G$ be a $P_6$-free graph, let $D$ denote a finite weight efficient dominating set (e.d.) of $G$, and for $v \in V$, let $D(v)$ denote an e.d. of $G$ containing $v$. Actually, one can choose a vertex $v$ of minimum degree $\delta(G)$ since by the e.d. property, either $v$ itself or a neighbor of $v$ dominates $v$. 

\medskip

Let $N_i(v)$ denote the $i$-th distance level of $v$, that is $N_i(v)=\{u \in V \mid d_G(u,v)=i\}$. Then, since $G$ is $P_6$-free, one has $N_i(v) = \emptyset$ for $i \geq 5$. By the e.d. property, one has 
\begin{equation}\label{N2Dvempty}
(N_1(v) \cup N_2(v)) \cap  D(v) = \emptyset. 
\end{equation}

Then, by the e.d. property, $D(v)$ is an e.d. of $G[\{v\} \cup N_2(v) \cup N_3(v) \cup N_4(v)]$ such that $N_2(v) \cap D(v) = \emptyset$. Moreover, without loss of generality, we can assume that $N_2(v)$ is a clique. 
Thus, the WED problem for $G$ is reduced to $\delta(G)$ cases of the WED problem for $G[N_2(v) \cup N_3(v) \cup N_4(v)]$ assuming that $v \in D(v)$ and 
$N_2(v) \cap D(v) = \emptyset$. 

\medskip
 
We call a vertex $x$ {\em $v$-forced} if $x \in D(v)$ for every e.d. $D(v)$ and {\em $v$-excluded} if $x \notin D(v)$ for every e.d. $D(v)$. Clearly, if $x$ is $v$-excluded then we can set $w(x)=\infty$, e.g., for all $x \in N_2(v)$, $w(x)=\infty$.   

\medskip

Let $Q_1, \ldots, Q_r$ denote the connected components of $G[N_3(v) \cup N_4(v)]$. By (\ref{N2Dvempty}), we have:
\begin{equation}\label{QiDvatleastone}
\mbox{ For each } i \in \{1,\ldots, r\}, \mbox{ one has } |Q_i \cap D(v)| \geq 1.
\end{equation}

Clearly, the $D(v)$-candidates in $Q_i$ must have finite weight.

\medskip

A component $Q_i$ is {\em trivial} if $|Q_i|=1$. Obviously, by (\ref{QiDvatleastone}), the vertices of the trivial components are in $D(v)$. Thus, trivial components consist of $v$-forced vertices. 

\medskip

Clearly, since $D(v)$ is an e.d. of finite weight, every $x \in N_2(v)$ must contact a component $Q_i$. 

\subsubsection{Join-reduction}

By the e.d. property of $D(v)$, condition (\ref{QiDvatleastone}) implies: 
\begin{equation}\label{atmostoneQijoin}
\mbox{ If there is } x \in N_2(v) \mbox{ with } x \join Q_i \mbox{ and } x \join Q_j, i \neq j, \mbox{ then } G \mbox{ has no e.d. } D(v).
\end{equation}

Thus, from now on, we can assume that every vertex $x \in N_2(v)$ has a join to at most one component $Q_i$. Moreover, if 
$x \join Q_i$ for some $i \in \{1,\ldots, r\}$ then for every neighbor $y \in Q_j$ of $x$, $j \neq i$, $y \notin D(v)$, i.e., $y$ is $v$-excluded. This means that we can set $w(y)=\infty$, and thus, $y \notin D(v)$ for any e.d. $D(v)$ of finite weight. 

\medskip

If $D(v)$ is an e.d., any vertex $x$ with $x \join Q_i$ for exactly one $i \in \{1,\ldots, r\}$ is correctly dominated if $|D(v) \cap Q_i|=1$. 
Since $D(v)$ is an e.d., $|D(v) \cap Q_i| \ge 2$ is impossible, that is, the $D(v)$-candidates in $Q_i$ are universal for $Q_i$; let $U_i$ denote the set of universal vertices in 
$Q_i$ (note that $U_i$ is a clique). Clearly, for $x \join Q_i$ we have:
\begin{equation}\label{Uiempty}
\mbox{ If  } U_i=\emptyset \mbox{ then } G  \mbox{ has no finite weight e.d. } D(v).
\end{equation}

Thus, for every $Q_i$ such that there is a vertex $x \in N_2$ with $x \join Q_i$, we can reduce $Q_i$ to the clique $U_i$, we can omit $x$ in $N_2$, and for every neighbor $y \in Q_j$ of $x$, $j \neq i$, we set $w(y)=\infty$.  

\medskip

For reducing WED on $G$ to WED on a unipolar graph $G'$, this is a first step which leads to the fact that finally, for every component $Q_j$ which is not a clique, every vertex in $N_2$ which contacts $Q_j$ also distinguishes $Q_j$. 
For this, we can do the following (since we use the subsequent algorithm a second time in Section \ref{sec:WEDP6frunipolargr} for WED on unipolar graphs, we slightly change the notions):   

\medskip

{\bf Join-Reduction Algorithm} 

\medskip

{\bf Given:} A graph $G=(A \cup B,E)$ such that $A_1,\ldots,A_k$ are the components of $G[A]$, and a vertex weight function $w$ 
with  $w(b)=\infty$ for all $b \in B$. \\

{\bf Task:} Reduce $G$ to graph $G'=(A' \cup B',E)$ with weight function $w'$ and components $A'_1,\ldots,A'_k$ of $G[A']$ such that for every $b \in B'$ and every $i \in \{1,\ldots,k\}$, if $A'_i$ is not a clique then $b$ distinguishes $A'_i$ if $b$ contacts $A'_i$, and $G$ has an e.d. with minimum finite weight if and only if $G'$ has such an e.d., or state that $G$ has no such e.d.

\medskip

{\bf begin}
\begin{itemize}  
\item[(a)] Determine the sets 
\begin{itemize}
\item[ ] $B_{join} := \{b \in B \mid$ there is an $i \in \{1,\ldots, k\}$ with $b \join A_i\}$ and 
\item[ ] $A_{join} := \{A_i \mid  i \in \{1,\ldots, k\}$ and there is a $b \in B$  with $b \join A_i \}$. 
\end{itemize}
\item[(b)] {\bf If} there is a vertex $b \in B_{join}$ and there are $i \neq j$ with $b \join A_i$ and $b \join A_j$ {\bf then} STOP $-$ $G$ has no e.d. of finite weight.\\
$\{$From now on, every $b \in B_{join}$ has a join to exactly one $A_i \in A_{join}$.$\}$

\item[(c)] {\bf For all} $b \in B_{join}$ and $A_i \in A_{join}$ such that $b \join A_i$ {\bf do}\\ 
{\bf begin}
\begin{itemize}
\item[(c.1)] Determine the set $U_i$ of universal vertices in $A_i$. {\bf If} $U_i=\emptyset$ {\bf then} STOP $-$ $G$ has no e.d. of finite weight {\bf else} set $A'_i:=U_i$.

\item[(c.2)] Set $w'(y) := \infty$ for every neighbor $y \in A \setminus A_i$ of $b$. 

\end{itemize}
{\bf end}

\item[(d)] For every $A_i \notin A_{join}$, set $A'_i:=A_i$, and finally set $A' := A'_1 \cup \ldots \cup A'_k$, $B':=B \setminus B_{join}$ and $G':=G[A' \cup B']$. 

\end{itemize}
{\bf end}

\medskip

For applying the Join-Reduction Algorithm to the distance levels of $v$, we set $B := N_2(v)$ and $A := N_3(v) \cup N_4(v)$.  

\begin{lemma}\label{WEDjoinredalg}
The Join-Reduction Algorithm is correct and can be done in time ${\cal O}(n^3)$. 
\end{lemma} 

{\bf Proof.}
{\em Correctness}: Assume that $D$ is a finite weight e.d. of $G$. Since for all $b \in B$, $w(b)=\infty$, we have $B \cap D = \emptyset$, and, as for (\ref{QiDvatleastone}), for every component $A_i$ we have $A_i \cap D \neq \emptyset$. 
If, as in (b), there is a vertex $b \in B_{join}$ such that $b$ has a join to (at least) two components of $G[A]$ then, as in (\ref{atmostoneQijoin}), by the e.d. property and since $A_i \cap D \neq \emptyset$ for every $i \in \{1,\ldots, k\}$, $G$ has no e.d. of finite weight, and we can stop. 

\medskip

Now, if $b \join A_i$ for exactly one $i \in \{1,\ldots, k\}$ then by the e.d. property, $b$ is dominated by exactly one $D$-vertex in $A_i$. Thus, if $|D \cap A_i|=1$, $b$ is correctly dominated but $D$ is no e.d. if $|D \cap A_i| \ge 2$. Thus, when reducing $B$ to $B'=B \setminus B_{join}$ as in (d), we have to ensure that $|D \cap A_i|=1$ which is done by considering the universal vertices $U_i$ of $A_i$ as described in (c.1); by (\ref{Uiempty}), if $U_i=\emptyset$ then we can stop.

\medskip

Moreover, clearly, for all neighbors $y$ of $b$ in $A_j$, $j \neq i$, $y \notin D$ which allows to set $w(y):=\infty$ as done in (c.2). 
 
\medskip

Clearly, every $A'_i$ which is not a clique has no join to any $b \in B_{join}$. Thus every such component is distinguished by any vertex $b \in B'$ contacting it. 

\medskip

{\em Time bound}:
For at most $n$ vertices $b \in B$, one has to check whether $b \join A_i$, $i \in \{1,\ldots, k\}$. For each of them, this can be done in  time ${\cal O}(n^2)$.
For each vertex $b$ with $b \join A_i$ for exactly one $i$, updating $w(y)=\infty$ for its neighbors $y \in A_j$ and adding $b$ to $A_i$ requires at most ${\cal O}(n)$ steps. Thus, the total time bound is  ${\cal O}(n^3)$. 
\qed

\subsubsection{Component-reduction}

In the next step, for reducing WED to unipolar graphs, we consider the components $Q_i$ which are not yet a clique. Without loss of generality, 
 after applying the Join-Reduction Algorithm to $G$ with $B=N_2(v)$ and $A=N_3(v) \cup N_4(v)$ (assuming that $v \in D(v)$), 
we can assume for $G$ that if $x \in N_2(v)$ has a neighbor in $Q_i$ then it has a neighbor and a non-neighbor in $Q_i$. For $1 \le i \le r$, let $Q^+_i(x)$ ($Q^-_i(x)$, respectively) denote the neighborhood (non-neighborhood, respectively) of $x$ in $Q_i$. Since $Q_i$ is a component of $G$, we have: If $x$ distinguishes $Q_i$ then it distinguishes an edge in $Q_i$. 

\medskip

For $x,x' \in N_2(v)$ and edges $y_1z_1$ in $Q_i$, $y_2z_2$ in $Q_j$, $i \neq j$, let $xy_1 \in E, xz_1 \notin E$ and 
$x'y_2 \in E, x'z_2 \notin E$. Then, since $G$ is $P_6$-free, we have:
\begin{equation}\label{P6edgecontact}
xy_2 \in E \mbox{ or } xz_2 \in E \mbox{ or } x'y_1 \in E \mbox{ or } x'z_1 \in E. 
\end{equation} 

Another useful $P_6$-freeness argument is the following:
\begin{equation}\label{Q+-join}
\mbox{For } x \in N_2(v) \mbox{ and } y \in Q^+_i(x), y \mbox{ does not distinguish any edge in } Q^-_i(x).  
\end{equation} 
 
{\em Proof.} If for $x \in N_2(v)$ and $y \in Q^+_i(x)$ with $xy \in E$, $y$ would distinguish an edge $z_1z_2 \in E$ in $Q^-_i(x)$ then for a common neighbor $u$ of $v$ and $x$, $\{v,u,x,y,z_1,z_2\}$ induce a $P_6$ in $G$ which is a contradiction.
$\diamond$

\medskip 
  
Let $B:=N_2(v)$; recall that without loss of generality, $B$ is a clique. We claim: 
\begin{equation}\label{univcontactexists}
\mbox{There is a vertex } b^* \in B \mbox{ which contacts } Q_i \mbox{ for every } i \in \{1,\ldots, r\}. 
\end{equation} 

{\em Proof.} Let $Q(b)$ denote the set of components $Q_i$ which are contacted (and thus, distinguished) by $b \in B$. Note that every component is contacted by at least one $b \in B$. By (\ref{P6edgecontact}) and since $G'$ is $P_6$-free, $Q(b)$ and $Q(b')$ for $b,b' \in B$, $b \neq b'$, cannot be incomparable. 
Thus either $Q(b) \subseteq Q(b')$ or vice versa, which implies that there is a vertex $b^* \in B$ with inclusion-maximal $Q(b^*)$, i.e., $b^*$ contacts every component $Q_i$. $\diamond$

\medskip

Let $q^* \in D(v)$ be the vertex dominating $b^*$; without loss of generality assume that $q^* \in Q_1$, and let $D(v,q^*)$ denote a finite weight e.d. containing $v$ and $q^*$. $Q_1$ is partitioned into 
\begin{itemize}
\item[ ] $Z:=N[q^*] \cap Q_1$, 
\item[ ] $W:=Q_1 \cap N(b^*) \setminus Z$, and 
\item[ ] $Y:=Q_1 \setminus (Z \cup W)$. 
\end{itemize}

Then clearly, the following properties hold:

\begin{lemma}\label{Q1partition}
\mbox{ }
\begin{itemize}
\item[$(i)$] $Z \cap D(v,q^*)=\{q^*\}$
\item[$(ii)$] $W \cap D(v,q^*)=\emptyset$
\item[$(iii)$] $Z \cojoin Y$ 
\item[$(iv)$] For every component $K$ of $G[Y]$, the set of $D(v,q^*)$-candidate vertices in $K$ is a clique.  
\end{itemize}
\end{lemma} 

{\bf Proof.}
$(i)$ and $(ii)$: Obviously hold by the e.d. property and the definition of $D(v,q^*)$.

\medskip

$(iii)$: By the definition of $Z$ and $Y$, we have $q^*y \notin E$ for every $y \in Y$. Let $K$ be a component in $G[Y]$. By the e.d. property, $K \cap D(v,q^*) \neq \emptyset$. Suppose to the contrary that there is an edge between some $z \in Z$ and $y \in K$. Then by the e.d. property, $y \notin D(v,q^*)$ and thus, there is $y^* \in K \cap D(v,q^*)$ with $y y^* \in E$ and $z y^* \notin E$ but now, $z$ distinguishes an edge in $K$ which is a contradiction to (\ref{Q+-join}) if $b^*z \in E$. Thus, we can assume that $b^*z \notin E$ but now, for a common neighbor $x$ of $v$ and $b^*$, $\{v,x,b^*,q^*,z,y\}$ induce a $P_6$ which is a contradiction.

\medskip

$(iv)$: By the e.d. property and by $(iii)$, for any component $K$ of $G[Y]$, $|K \cap D(v,q^*)| \ge 1$.
By $(ii)$ and the definition of $W$ and $Y$, $K$ must have a neighbor in $W$. By (\ref{Q+-join}), every neighbor of $K$ in $W$ has a join to $K$, thus $|K \cap D(v,q^*)| = 1$ and we can reduce $K$ to a clique (only the universal vertices of $K$ are the $D(v,q^*)$-candidate vertices); if there is no such universal vertex then $G$ has no e.d. $D(v,q^*)$.  
\qed

\medskip

For the algorithmic approach, we set $w(y)=\infty$ for every $y \in W$ and for every non-universal vertex $y \in K$ in any component $K$ of $G[Y]$. 

\medskip

For $i \ge 2$, let $Q^+_i$ ($Q^-_i$, respectively) denote the neighborhood (non-neighborhood, respectively) of $b^*$ in $Q_i$.  

Clearly, by the e.d. property, for every $i \ge 2$, $Q^+_i \cap D(v,q^*)=\emptyset$; set $w(y)=\infty$ for every $y \in Q^+_i$. Thus, the components of $G[Q^-_i \setminus Q^+_i]$ must contain the corresponding $D(v,q^*)$-vertices. 
Again, as in Lemma \ref{Q1partition} $(iv)$, for each such component $K$, the $D(v,q^*)$-candidates must be universal vertices for $K$ since by (\ref{Q+-join}), two such $D(v,q^*)$-candidates in $K$ would have a common neighbor in $Q^+_i$, i.e., only the universal vertices of component $K$ are the $D(v,q^*)$-candidate vertices; set $w(y)=\infty$ for every non-universal vertex $y \in K$.  

\medskip

Thus, for every potential $D(v)$-neighbor $q^*$ of $b^*$, we can reduce the WED problem for $G$ to the WED problem for $G'$ consisting of the clique $B$ and the $P_3$-free subgraph induced by the corresponding cliques of universal vertices in components $K$. 

\medskip

Clearly, the $D(v,q^*)$-candidates in the cliques of the $P_3$-free subgraph can be chosen corresponding to optimal weights. 

\medskip

Summarizing the facts given by Lemma \ref{Q1partition} and the paragraph after the proof of Lemma \ref{Q1partition} for $i \ge 2$, we can do the following: 

\medskip

{\bf Component-Reduction Algorithm} 

\medskip

{\bf Given:} A subgraph $\hat{G}=G[N_2(v) \cup N_3(v) \cup N_4(v)]$ such that $K_1,\ldots,K_s$ are the clique components and $Q_1,\ldots,Q_r$ are the non-clique components of $G[N_3(v) \cup N_4(v)]$. Moreover, for all $b \in N_2(v)$, $w(b)=\infty$ and when $b$ contacts $Q_i$ then it has a neighbor and a non-neighbor in $Q_i$. \\
{\bf Task:} Reduce $\hat{G}$ to at most $n$ graphs $G'(q^*)$ such that every component of $G'(q^*)$ in $N_3(v) \cup N_4(v)$ is a clique in $G$ (and thus, $G'(q^*)$ is unipolar) and $G$ has an e.d. with finite weight if and only if there is a $q^*$ such that $G'(q^*)$ has such an e.d. 

\medskip

{\bf begin}
\begin{itemize}
\item[(a)] Determine a vertex $b^* \in N_2(v)$ contacting every $Q_i$, $i \in \{1,\ldots, r\}$.   

\item[(b)] For every $q^* \in N(b^*)$ with $w(q^*) < \infty$, reduce $Q_i$ according to Lemma \ref{Q1partition} and the paragraph after the proof of Lemma \ref{Q1partition} for $i \ge 2$ such that finally, $G'(q^*)$ is unipolar. 

\end{itemize}
{\bf end}

\medskip

Since for every $q^* \in N(b^*)$, $G'(q^*)$ is a unipolar graph and starting with $G$, the join-reduction and the component-reduction phase can be done in polynomial time and lead to at most $n^2$ such unipolar graphs, we have shown: 

\begin{lemma}\label{WEDredunipolargr}
If WED is solvable in polynomial time on $P_6$-free unipolar graphs then it is solvable in polynomial time on $P_6$-free graphs.
\end{lemma} 

A more exact time bound will be given in Section \ref{algP6fr}.

\subsection{Solving WED on $P_6$-free unipolar graphs in polynomial time}\label{sec:WEDP6frunipolargr}

The key result of this subsection is the following:
\begin{lemma}\label{WEDP6frunipolargr}
Let $G=(V,E)$ be a connected graph, where $V$ admits a partition $V = A \cup B$ such that:
\begin{itemize}
\item[$(i)$] $G[A]$ is the disjoint union of cliques $A_1,\ldots,A_k$ and $G[B]$ is a complete subgraph.
\item[$(ii)$] If for distinct $b_1,b_2 \in B$, $b_1$ distinguishes an edge $x_1x_2$ in $A_i$ and $b_2$ distinguishes an edge $y_1y_2$ in $A_j$, $i \neq j$, then either $b_2$ contacts $\{x_1,x_2\}$ or $b_1$ contacts $\{y_1,y_2\}$.
\end{itemize}
Then it can be checked in polynomial time whether $G$ has a finite weight e.d. $D$ with $B \cap D = \emptyset$.
\end{lemma}

Clearly, for a $P_6$-free unipolar graph, condition $(ii)$ of Lemma \ref{WEDP6frunipolargr} is fulfilled. For proving Lemma \ref{WEDP6frunipolargr}, we subsequently collect various propositions.

\medskip

As a first step, we again reduce $G$ corresponding to the Join-Reduction Algorithm of Section \ref{unipolarred}: Since $B \cap D=\emptyset$, clearly, $|D \cap A_i|=1$ for every 
$i \in \{1,\ldots,k\}$. Thus, if $A_i=\{a_i\}$ then $a_i$ is a forced $D$-vertex; from now on, we can assume that every $A_i$ is nontrivial. 

\medskip

Moreover, every $b \in B$ must contact at least one $A_i$, and if $b$ has a join to two components $A_i,A_j$, $i \neq j$, then $G$ has no e.d. Thus, by (\ref{atmostoneQijoin}) and the subsequent paragraph in Section \ref{unipolarred}, from now on, we can assume that no vertex $b \in B$ has a join to any $A_i$, i.e., if $b$ contacts $A_i$ then it distinguishes $A_i$.     

\medskip

 Again, as by (\ref{univcontactexists}), there is a vertex $b^* \in B$ which contacts every $A_i$. However, we need a stronger property - for this, we define the following notions: 
\begin{definition}\label{overtakedef}
For vertices $b_1,b_2 \in B$ and a nontrivial component $K=A_i$ of $A$, we say:
\begin{itemize}
\item[$(i)$] {\em $b_2$ overtakes $b_1$ for $K$} if $b_2$ distinguishes an edge in $K \setminus N(b_1)$.
\item[$(ii)$] {\em $b_2$ includes $b_1$ for $K$} if $N(b_2) \cap K \supseteq N(b_1) \cap K$.
\item[$(iii)$] {\em $b_2$ strictly includes $b_1$ for $K$} if $N(b_2) \cap K \supset N(b_1) \cap K$.
\item[$(iv)$] {\em $b_1$ and $b_2$ cover $K$} if $N(b_1) \cup N(b_2) = K$.
\item[$(v)$] {\em $b_1 \rightarrow b_2$} if $b_2$ overtakes $b_1$ for at least three distinct nontrivial components of~$A$. 
\item[$(vi)$] $b^* \in B$ is a {\em good vertex} of $B$ if for none of the vertices $b \in B \setminus \{b^*\}$, $b^* \rightarrow b$ holds.  
\end{itemize}
\end{definition}

Let $H=(B,\vec{F})$ denote the directed graph with vertex set $B$ and edges $b \rightarrow b' \in \vec{F}$ as defined in $(v)$. Thus, a good vertex of $B$ is one with outdegree 0 with respect to $H$. As usual, $H$ is a {\em directed acyclic graph} ({\em dag}) if there is no directed cycle in $H$. It is well known that any dag has a vertex with outdegree 0. 

\medskip

The following is easy to see by using Condition $(ii)$ of Lemma \ref{WEDP6frunipolargr} and the e.d. property: 
 
\begin{claim}\label{overtakeprop}
For vertices $b_1,b_2 \in B$, we have: 
\begin{itemize}
\item[$(i)$] $b_1$ and $b_2$ cover at most two $A_i$, $A_j$, $i,j \in \{1,\ldots,k\}$, $i \neq j$.

\item[$(ii)$] If $b_2$ overtakes $b_1$ for $A_i$ then $b_1$ does not overtake $b_2$ for $A_j$, $j \neq i$. 

\item[$(iii)$] If $b_2$ overtakes $b_1$ for some $A_i$ then, since $b_1$ and $b_2$ cover at most two $A_j,A_{\ell}$, $j,\ell \neq i$,
$b_2$ includes $b_1$ for the remaining components $A_i$, $i \neq j,\ell$.

\item[$(iv)$] If $b_2$ overtakes $b_1$ for some $A_i, A_j$, $i \neq j$, then $b_2$ strictly includes $b_1$ for $A_i, A_j$.

\end{itemize}
\end{claim}
 
\begin{claim}\label{Hdag}
$H$ is a directed acyclic graph.  
\end{claim} 
 
{\bf Proof.} 
We claim that there is no directed cycle in $H$ with vertices $b_1,\ldots,b_h \in B$ such that $b_1 \rightarrow b_2 \rightarrow \ldots \rightarrow b_h \rightarrow b_1$. Assume to the contrary that such a directed cycle with vertices $b_1,\ldots,b_h \in B$ exists.

\medskip

If $h = 2$, then since $b_1 \rightarrow b_2$ one has that $b_2$ includes $b_1$ for all $A_i$ except for at most two such members by Claim \ref{overtakeprop} $(iii)$ and $(iv)$. This implies that $b_2 \not \rightarrow b_{1}$, and thus, there is no cycle $b_1 \rightarrow b_2 \rightarrow b_1$.

\medskip

If $h \geq 3$, then we first show:  

\begin{itemize}
\item[ ] $(*)$ $b_h$ includes $b_{h-2}$ for all $A_i$ except for at most two of them. 
\end{itemize}

{\em Proof of} $(*)$. Since $b_{h-1} \rightarrow b_h$, one has that $b_h$ includes $b_{h-1}$ for all 
$A_i$ except for at most two of them by Claim \ref{overtakeprop} $(iii)$ and $(iv)$.

On the other hand, since $b_{h-2} \rightarrow b_{h-1}$, one has that $b_{h-1}$ includes $b_{h-2}$ for all $A_i$ except for at most two of them by Claim \ref{overtakeprop} $(iii)$ and $(iv)$; in particular $b_{h-1}$ strictly includes $b_{h-2}$ for three members of $A_i$ by definition of relation $\rightarrow$
and by Claim \ref{overtakeprop} $(iv)$. This implies that $b_h$ strictly includes $b_{h-2}$ for at least one $A_i$. Then, since $b_h$ distinguishes $A_i$, $b_h$ overtakes $b_{h-2}$ for $A_i$. 

Then, by Claim \ref{overtakeprop} $(iii)$ and since $b_h$ (strictly) includes $b_{h-2}$ for $A_i$, one has that: $b_h$ and $b_{h-2}$ cover at most two of $A_1,\ldots,A_k$ and $b_h$ includes $b_{h-2}$ for the remaining ones. Thus, the assertion $(*)$ is proved.

\medskip

Then let us show that more generally, $b_h$ includes $b_{j}$, for $j \in \{h-2, \ldots, 1\}$, for all $A_i$ except for at most two of them. That can be proved consecutively for $j = h-2, \ldots, 1$, as shown in the previous paragraph (in particular, for $j = h-3,\ldots,1$, the starting point is that $b_h$ includes $b_{j+1}$ for all $A_i$ except for at most two of them). That is, one finally has that $b_h$ includes $b_{1}$ for all $A_i$ except for at most two of them. This implies that $b_h \not \rightarrow b_{1}$ which is a contradiction.

\medskip

Thus there are no vertices $b_1,\ldots,b_h \in B$ such that $b_1 \rightarrow b_2 \rightarrow \ldots \rightarrow b_h \rightarrow b_1$, i.e., $H$ is a dag which proves Claim \ref{Hdag}. 
\qed
 
\medskip

As already mentioned, it is well known that every dag has a vertex of outdegree 0. Thus, Claim \ref{Hdag} implies: 
 
\begin{claim}\label{goodvex}
There is a good vertex $b^* \in B$. 
\end{claim} 

Let $b^*$ be such a good vertex. Then, since by the condition in Lemma \ref{WEDP6frunipolargr}, $B \cap D = \emptyset$, $b^*$ must have a $D$-neighbor $a^* \in A \cap N(b^*) \cap D$; this has to be iterated for all vertices in $A \cap N(b^*)$. Let $D(a^*)$ denote an e.d. with $a^* \in D(a^*)$; without loss of generality, assume that $a^* \in A_1$. Clearly, $(A_1 \setminus \{a^*\}) \cap D(a^*) = \emptyset$. Without loss of generality, let us assume that $A_1 = \{a^*\}$.
Since $a^*$ dominates $b^*$, each neighbor of $b^*$ in $A_i$, $i \ge 2$, is not in $D(a^*)$. For $i  \in \{2,\ldots,k\}$, let $A'_i := A_i \setminus N(b^*)$, and let $A' = \{a^*\} \cup A'_2 \cup \ldots \cup A'_k$. Obviously, one has: 

\begin{enumerate}
\item[(a)] For each $A'_i$, $|A'_i \cap D(a^*)| = 1$.
\end{enumerate}

Moreover, as before, we can assume:  
\begin{enumerate}
\item[(b)] For each vertex $b \in B$, $b$ does not have a join to two distinct $A'_i$, $A'_j$.
\item[(c)] If vertex $b \in B$ has a join to exactly one $A'_i$ then it does not contact the remaining components $A'_j$, $j \neq i$.
\end{enumerate}

Next we claim: 
\begin{enumerate}
\item[(d)] At most three distinct components $A'_i,A'_j,A'_{\ell}$ are distinguished by some vertex of $B \setminus \{b^*\}$.
\end{enumerate}

{\em Proof.} Assume to the contrary that four of $A'_1,\ldots,A'_k$ are distinguished by some vertex of $B \setminus \{b^*\}$. Then, similarly as for (\ref{univcontactexists}), it is easy to see that there is a vertex $b \in B \setminus \{b^*\}$ contacting all such four components $A'_i$ of $A'$. Then by propositions (b) and (c), vertex $b$ distinguishes at least three such components, that is $b^* \rightarrow b$ which is a contradiction to the assumption that $b^*$ is a good vertex. $\diamond$

\medskip

Summarizing, by the above, $D(a^*)$ exists if and only if 
\begin{itemize}
\item[$(i)$] the above properties hold and
\item[$(ii)$] $G[A' \cup B]$ has the (weighted) e.d. $D(a^*)$ with $B \cap D(a^*) = \emptyset$. 
\end{itemize}

Checking $(i)$ can be done polynomial time (actually one should check just if some of the above properties hold). Checking $(ii)$ can be done in polynomial time as shown below: For the components of $G[A']$, let 
\begin{itemize}
\item[ ] $C_{1}(A')$ be the set of those components of $G[A']$ which are not distinguished by any vertex of $B$, and  
\item[ ] $C_{2}(A')$ be the set of those components of $G[A']$ which are distinguished by some vertex of $B$. 
\end{itemize}

For each member $K$ of $C_{1}(A')$, one can select one vertex of $K$ as a candidate for $D(a^*) \cap K$, since such vertices form a clique and have respectively the same neighbors in $G[(A' \cup B) \setminus K]$ (for WED, one can select a vertex of minimum weight). 

\medskip

Concerning $C_{2}(A')$, one has $|C_{2}(A')| \leq 3$ by property (d).
Then the set $\{(a^*,a_2,\ldots,a_k): a_i \in A'_i, i \in \{2,\ldots,k\}\}$ of $k$-tuples of candidate vertices in $D(a^*)$ contains $O(n^3)$ members by property (d).
Thus one can check in polynomial time if $D(a^*)$ exists.
This completes the proof of Lemma \ref{WEDP6frunipolargr}.

\medskip

By combining Lemmas \ref{WEDredunipolargr} and \ref{WEDP6frunipolargr}, we obtain: 

\begin{theorem}\label{WEDP6frPol}
WED is solvable in polynomial time for $P_6$-free graphs.  
\end{theorem}  

\subsection{The algorithm for WED on $P_6$-free graphs}\label{algP6fr}

\subsubsection{WED for $P_6$-free unipolar graphs}

{\bf Algorithm WED for $P_6$-free unipolar graphs}

\medskip

{\bf Given:} A $P_6$-free unipolar graph $G=(A \cup B,E)$ such that $B$ is a clique and $G[A]$ is the disjoint union of cliques $A_1,\ldots,A_k$.\\
{\bf Task:} Determine an e.d. of $G$ with minimum finite weight if there is one or state that $G$ has no such e.d.

\begin{itemize}
\item[(a)] Reduce $G$ to $G'$ as in the Join-Reduction Algorithm. $\{$From now on, we can assume that for every $b \in B$ and every $i \in \{1,\ldots,k\}$, $b$ distinguishes $A_i$ if $b$ contacts $A_i$.$\}$  

\item[(b)] Construct the dag $H$ according to Definition \ref{overtakedef} $(v)$, and determine a good vertex $b^* \in B$ in $H$. 

\item[(c)] For every neighbor $a^* \in A'$ of $b^*$, determine the ${\cal O}(n^3)$ possible tuples of $D(a^*)$-candidates and check whether they are an e.d. of finite weight.  

\item[(d)] Finally, choose an e.d. of minimum finite weight or state that $G'$ does not have such an e.d.

\end{itemize}

\begin{theorem}\label{EDP6frunipolargrpolcorr} 
Algorithm WED for $P_6$-free unipolar graphs is correct and can be done in time ${\cal O}(n^4 m)$. 
\end{theorem} 

{\bf Proof.}
{\em Correctness:} The correctness of Algorithm WED for $P_6$-free unipolar graphs follows from the reduction arguments described in Section \ref{unipolarred} 
and the arguments for $P_6$-free unipolar graphs described in Section \ref{sec:WEDP6frunipolargr}.

\medskip

{\em Time bound:} 
Step (a) can be done in time ${\cal O}(n^3)$ by Lemma \ref{WEDjoinredalg}.
Step (b) can be done in time ${\cal O}(n^4)$: Constructing $H$ can be done in time ${\cal O}(n^4)$ since there are at most ${\cal O}(n^2)$ edges $b \rightarrow b'$ in $H$, and checking for each pair $(b,b')$ whether $b \rightarrow b'$ takes ${\cal O}(n^2)$ time. 
Step (c) can be done in time ${\cal O}(n^4 m)$: For at most $n$ possible neighbors $a^* \in A'$ of $b^*$, check in linear time for at most ${\cal O}(n^3)$ possible tuples of $D(a^*)$-candidates whether they are an e.d. of finite weight. 
Obviously, Step (d) can be done in linear time. 
\qed

\subsubsection{WED for $P_6$-free graphs}

{\bf Algorithm WED for $P_6$-free graphs}

\medskip

{\bf Given:} A $P_6$-free graph $G=(V,E)$.\\
{\bf Task:} Determine an e.d. of $G$ with minimum finite weight if there is one or state that $G$ has no such e.d.

\medskip

Determine a vertex $v_0$ of minimum degree $\delta(G)$, and do the following for every $v \in N[v_0]$:
\begin{itemize}
\item[(a)] Determine the distance levels $N_i(v)$, $1 \le i \le 4$. 

\item[(b)] For $B=N_2(v)$ and $A=N_3(v) \cup N_4(v)$, reduce $G$ to $G'$ as in the Join-Reduction Algorithm. $\{$From now on, we can assume that for every $b \in B$ and every $i \in \{1,\ldots,k\}$, $b$ distinguishes $A_i$ if $b$ contacts $A_i$.$\}$  

\item[(c)] According to the Component-Reduction Algorithm, determine a vertex $b^* \in B$ contacting every component in $G[A]$ which is not a clique, and for  
every neighbor $q^* \in N(b^*) \cap A$, do:

\begin{itemize}
\item[(c.1)] Reduce $G$ to $G'(v,q^*)$ by the Component-Reduction Algorithm. $\{$Now, $G'(v,q^*)$ is $P_6$-free unipolar.$\}$   

\item[(c.2)] Carry out the Algorithm WED for $P_6$-free unipolar graphs for input $G'(v,q^*)$ with its weight function.
\end{itemize}

\item[(d)] Finally, for every resulting e.d. set, check whether it is indeed a finite weight e.d. of $G$,   
choose an e.d. of minimum finite weight of $G$ or state that $G$ does not have such an e.d.

\end{itemize}

\begin{theorem}\label{EDP6frgrpolcorr} 
Algorithm WED for $P_6$-free graphs is correct and can be done in time ${\cal O}(\delta(G) n^5 m)$. 
\end{theorem} 

{\bf Proof.}
{\em Correctness:} The correctness of Algorithm WED for $P_6$-free graphs follows from the reduction arguments described in Section \ref{unipolarred} 
and the correctness of the Algorithm WED for $P_6$-free unipolar graphs.

\medskip

{\em Time bound:} Since, as a starting point, we choose a vertex $v_0$ of minimum degree $\delta(G)$, Steps (a)-(c) have to be done at most $\delta(G) \le n$ times. 
Obviously, Step (a) can be done in linear time ${\cal O}(n + m)$.  
Step (b) can be done in time ${\cal O}(n^3)$ by Lemma \ref{WEDjoinredalg}. 
Obviously, determining a vertex $b^* \in B$ contacting every component in $G[A]$ which is not a clique, can be done in time ${\cal O}(n^3)$. 
Steps (c.1) and (c.2) have to be done at most $n$ times. By Theorem \ref{EDP6frunipolargrpolcorr}, Step (c.2) can be done in time ${\cal O}(n^4 m)$, and this is an upper bound for Step (c.1) as well.
Finally, Step (d) can be done in linear time. 
Thus, Algorithm WED for $P_6$-free graphs can be done in time ${\cal O}(\delta(G) n^5 m)$ (which is at most ${\cal O}(n^6 m)$). 
\qed

\section{Conclusion}

As mentioned, the direct approach for solving WED on $P_6$-free graphs gives a dichotomy result for the complexity of WED on $F$-free graphs. In \cite{BraEscFri2015}, using an approach via $G^2$, it was shown that WED can be solved in polynomial time for $P_6$-free chordal graphs, and a conjecture in \cite{BraEscFri2015} says that for $P_6$-free graphs with e.d., the square is perfect which would also lead to a polynomial time algorithm for WED on $P_6$-free graphs but the time bound of our direct approach is better than in the case when the conjecture would be true.    

\begin{footnotesize}

\end{footnotesize}

\end{document}